# SOFTWARE AS A SERVICE – COMMON SERVICE BUS *(SAAS-CSB)*

R. Swaminathan[1], K. Karnavel[2]

[1]Assistant Professor, Department of Computer Science & Engineering, Sakthi Engineering College, Chennai
`swaminathan1984@gmail.com`

[2]Assistant Professor, Department of Computer Science & Engineering, Anand Institute of Higher Technology, Chennai
`treseofkarnavel@gmail.com`

*ABSTRACT*

*Software-as-a-Service (SaaS) is a form of cloud computing that relieves the user from the concern of hardware, software installation and management. It is an emerging business model that delivers software applications to the users through Web-based technology. Software vendors have varying requirements and SaaS applications most typically support such requirements. The various applications used by unique customers in a single instance are known as Multi-Tenancy. There would be a delay in service when the user sends the data from multiple applications to multiple destinations and from multiple applications to single destination due to the use of single CSB. This problem can be overcome by using multiple CSB concepts and hence multiple senders can efficiently send their data to multiple receivers at the same time. The multiple clouds are monitored and managed by the SaaS-CSB portal. The idea of SaaS-CSB Portal is to provide a single pane of glass for the user to consume and govern any service from any cloud. Thus, SaaS-CSB application allows companies to save their IT cost and valuable time.*

*KEYWORDS*

cloud computing, SaaS, Multi-Tenancy, CSB and SaaS-CSB portal.

## 1. INTRODUCTION

The Common Service Bus (CSB) acts as the middleware between applications that addresses the fundamental need for application integration. Single-tenant CSB cannot support multiple clients simultaneously and hence we focus on making CSB's multi-tenant aware. Multi-tenancy are the key enablers that allow Cloud computing solutions to serve multiple customers from a single instance. In Short, *"Multi-tenancy is an architecture in which a single instance of a software application serves multiple customers and each customer is called a tenant"*[1][2]. From our findings, multiple producers will upload their own application, and these applications are transferred to the multiple CSB through agents, which are uploaded into the database. From this database, consumer can consume their required application. The applications in multiple CSB's are monitored, consumed and managed using SaaS-CSB portal.

## 2. EXISTING SYSTEM

The existing system uses only single CSB in SaaS environment, where the data are sent through single CSB which lead to delay in data delivery[1]. The process exists between single application to single destination and from single application to multiple destinations, which leads to poor performance, where the data may not be sent efficiently[2].

### 2.1 Drawbacks

1. It does not ensure the efficient data delivery to the receiver, as it deals with single tenant.
2. Congestion occurs as the data are overcrowded in the queue.
3. There may be loss of data due to congestion.

## 3. PROPOSED SYSTEM

The proposed system is about creating the multiple CSBs concept in SaaS, which is based on Multi-tenancy. By this, the data is delivered to all consumers rapidly without any congestion. This leads to the high performance of data delivery to the consumers. This application is embedded in the cloud and it has been consumed, monitored and managed by the SaaS-CSB Portal[4].

## 4. ARCHITECTURE

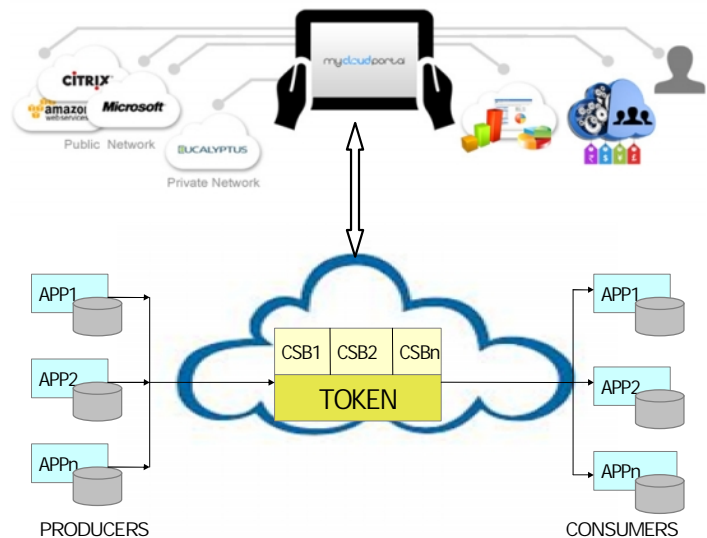

Fig:1 SaaS-CSB Architecture Diagram

# 5. MODULES

## 5.1 Common Service Bus (CSB)

### 5.1.1 Producer

The task of the producer is to upload their application into the CSB through agents. In our project we are uploading the bank and medical applications. Producer1 will enroll into the bank registration form and upload their data. Similarly producer2 will enroll into the medical registration form and upload their data [9].

### 5.1.2 Token

The purpose of the token is to manage the data that are available in the CSB. In our work, the bank application available in one CSB is managed and configured by the token and the similar work is done for the medical application also. The tokens are generated when an application sent by the producer is passed into the CSB[3].

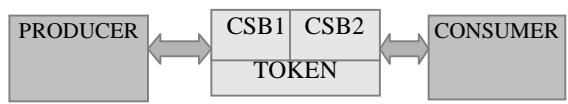

Fig:2 Token Generation

### 5.1.3 Consumer

The data from the CSB's are then stored in the database so that it will be available for future references. In our project consumer1 will consume the bank details from the database1 and similarly consumer2 will consume the medical details from the database2.

## 5.2 Saas-CSB Portal

### 5.2.1 Connect and Manage

Connect and manage acts as a cloud service gateway or a cloud broker in the Saas-CSB Portal in order to help the application provider to create and manage infrastructure services like Compute, Volume, Snapshot, KeyPair, Security Group & IP Address from IaaS cloud[5].

### 5.2.2 User and Access control

It uses the cloud governance features to define role based access controls and security at the Manager, Admin and User levels using the cloud control panel. This provides the security to all, who are being connected with the SaaS-CSB Portal.

### 5.2.3 Usage and Cost Report

It is used to generate project, department, account or user based reports by using the cloud management features. It displays the overall cost and information about the resources. Based on the usage the cost has been allocated based on pay-as-you-go basis[5].

### 5.2.4 Dashboard

Dashboard provides the overall view of the resource running in the cloud. As a cloud management solution, we can get the view of the cloud's health, workflows, cost and usage based on predefined roles. It displays the details of the cloud in a single panel.

### 5.2.5 Workflow

It uses the cloud governance features to setup workflows which mirror the business processes and control the internal cloud consumption by approving or rejecting infrastructure requests by the users. It explains the internal process flow, on how the cloud is being connected with other services[6][10].

### 5.2.6 Product Catalog

Product Catalog is used to setup the price of the package that we use in our infrastructure products in local currencies. The currency can be assigned in rupees, dollars, euro, etc. Based on the country where the SaaS-CSB Portal is used[7][8].

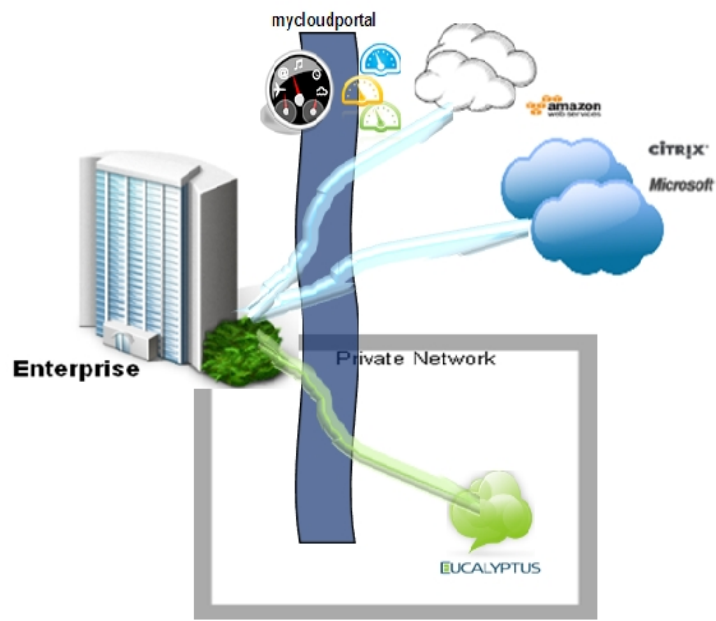

Fig:3 SaaS-CSB Portal

### 5.2.7 Advantages

- Faster communication
- No congestion problem
- No Loss of data
- Efficient
- High performance

## 6. COMPARISON

In the existing system, the redundancy of the data is 0%.Whereas in the proposed system the redundancy of the data is 100% . The scalability and performance ratio is increased from 30% to 70%. Additionally, in the proposed system, the operational cost is reduced to 20%, whereas in the existing system, operational cost was 90%.

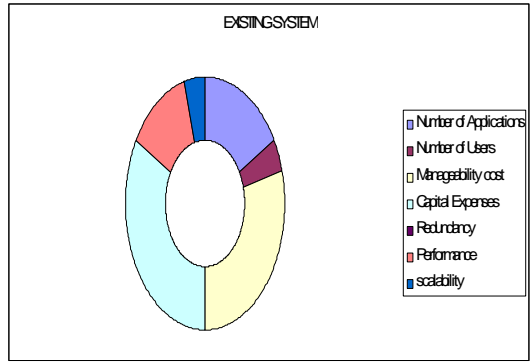

Fig:4 Pie chart for Existing System

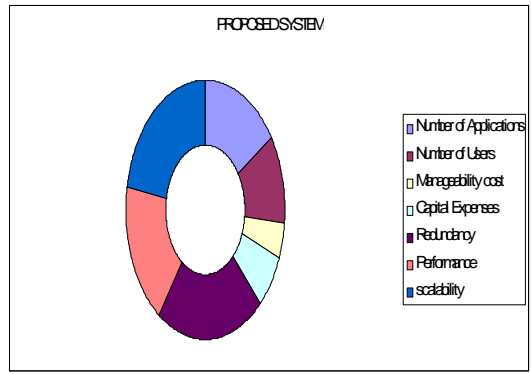

Fig:4 Pie chart for Proposed System

## 7. CONCLUSION AND FUTURE WORKS

In the previous sections, we proposed an approach for creating CSB multi-tenant aware, which would be able to serve multiple consumers from a multiple producers simultaneously in a SaaS environment. Here the token is used to manage the data that are available in the CSB. This application is embedded in a cloud which is managed by a SaaS-CSB Portal. This portal is used to consume, monitor and manage the clouds.

The enhancement of the project of creating CSB multi-tenant in PaaS and in IaaS environment that enable multiple producer to serve multiple consumer simultaneously in these (PaaS, IaaS) environments with high security. We can further enhance the SaaS-CSB Portal by metering, monitoring, billing the cloud together with customer engagement.